\documentclass[a4paper]{article}

\usepackage{graphicx}
\begin{document}

\title{ Dynamics of aeolian sand ripples}

\author{ %\lineskip 1em
Zolt\'an Csah\'ok$^{1,3}$, Chaouqi Misbah$^1$,\\
Fran\c{c}ois Rioual$^2$, and Alexandre Valance$^2$\\[1em]
$^1$ {\sl
Laboratoire de Spectrom\'etrie Physique, Universit\'e Joseph Fourier (CNRS)}\\
{\sl Grenoble I, B.P. 87, Saint-Martin d'H\`eres, 38402 Cedex, France}\\[0.5em]
$^2$ {\sl Groupe Mati\`ere Condens\'ee et Mat\'eriaux, UMR 6626,}\\
{\sl Universit\'e Rennes 1, 35042 Rennes Cedex, France }\\[0.5em]
$^3$ {\sl MTA Res. Inst.  for Technical Physics and Materials Science,}\\
{\sl POBox 49, H-1525 Budapest, Hungary}
}
% \date{\today}
\maketitle

\begin{abstract}
We analyze theoretically the dynamics of aeolian sand ripples.  In order
to put the study in the context we first review existing models.
%We  argue on the local character of sand ripple formation.
%Using a hydrodynamical model we derive a nonlinear equation for the sand
%profile. 
This paper is a continuation of two
previous papers\cite{Csahok99a,Valance99}, the first
one is based on symmetries and the second
on a hydrodynamical model. 
We  show how the hydrodynamical
model may be modified to recover the missing terms
that are dictated by symmetries.
The symmetry and conservation arguments are powerful in that 
the form of the equation is model-independent.
We then present
an extensive numerical and analytical
analysis of the generic sand ripple equation. We find
that at the initial stage the wavelength of the ripple is that
corresponding to the linearly most dangerous mode. At later
stages the profile undergoes a coarsening process leading 
to a significant increase of the wavelength. We find
that including the next  higher order nonlinear term
in the equation, leads naturally to a saturation of the local slope.
We analyze both
analytically and numerically the coarsening stage, in terms of
a dynamical exponent for the mean wavelength increase. We discuss
some  future lines of investigations.

\end{abstract}

\noindent PACS numbers: 83.70.Fn,81.05.Rm,47.20.-k

%11111111111111111111111111111111111111111111111111111111111111111111111
\newcommand{\beqa}{\begin{eqnarray}}
\newcommand{\beq}{\begin{equation}}
\newcommand{\eeqa}{\end{eqnarray}}
\newcommand{\eeq}{\end{equation}}
\newcommand{\dbyd}[2]{{\partial{#1}\over\partial{#2}}}
\newcommand{\dbydd}[2]{{\partial^2{#1}\over\partial{#2}}}
\newcommand{\dbydfl}[2]{{\partial{#1}/\partial{#2}}}
\newcommand{\kom}{\>,}
\newcommand{\pnt}{\>.}

\section{Introduction}

Perhaps the most ancient and fascinating out-of-equilibrium example
of spontaneous pattern formation
known in nature is that exhibited
by a sand bed subjected to wind. If wind is   strong enough (but not too
strong to prevent erosion), of the order of few m/s,
sand grains enter into a perpetual motion 
causing ultimately the sand bed to become unstable to 
ripple formation, commonly referred to as {\it aeolian sand ripples}. 
The typical wavelength is of the order  of few cm (in some deserts,
in Libya, however, ripples continue to coarsen leading to wavelengths which are
much larger -- several m --,
and are usually called ridges).
Geologists, in particular, have been intrigued that
such an apparently simple system as sand turns from an intially
structureless state into a rather organized structure
in a quite robust and reproducible fashion, despite the turbulent
air flow that causes the ripple formation. Following the seminal
work of Bagnold~\cite{Bagnold41} many researchers have achieved a significant
contribution to the understanding of ripple formation both experimentally
and theoretically. This field of research has known more recently
an upsurge of interest  as a part of the puzzling behaviour of granular
media. Despite the fact that sand is a very familiar material,
the understanding of its static
and dynamical properties still poses a formidable challenge
to theoretical modelling. Unlike elastic, viscoelastic materials, and 
Newtonian fluids, there is yet no universal continuum theory (such that
leading to  the Lam\'e
or Navier-Stokes equations) to describe in a effective manner
the behaviour of  granular media. A major difficulty, in our opinion,
lies in the broad spectrum
of length and time scales.  Despite this  situation, various
tools have been used to describe in a more or less ad hoc way granular
media. With regard to ripple formation, we may cite
(i) molecular dynamics introducing
empirical laws of collision,
(ii) Monte Carlo simulations, trying 
to mimic what is our feeling about rules of collision and rearrangements,
(iii)
hydrodynamical theories inspired from Bagnold's view.
Concerning
the birth of ripples the view of Bagnold is largely adopted
and it will be reviewed shortly here. 

\begin{figure}[h]
\begin{center}
\includegraphics{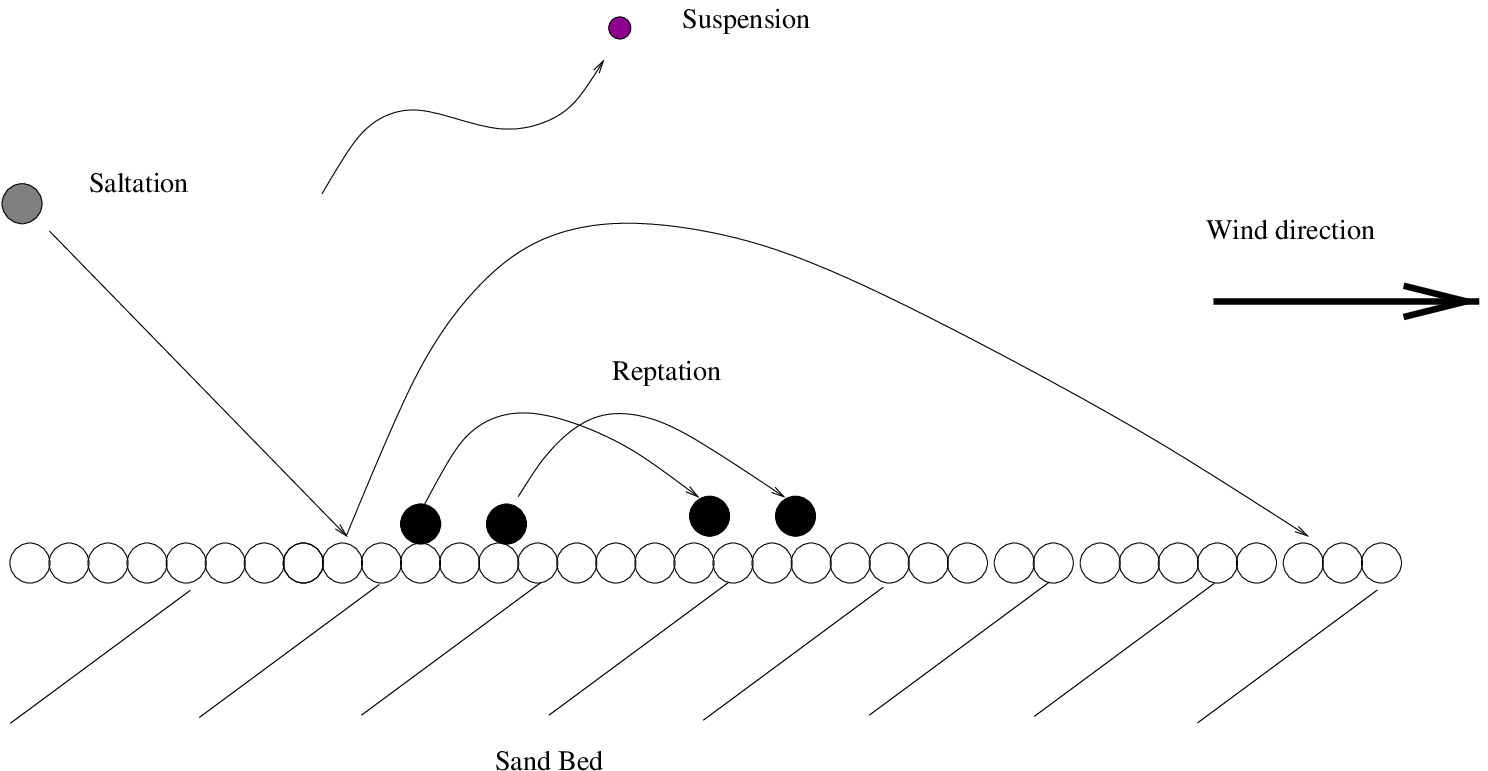}
\end{center}
\caption{Different mechanisms of transport of sand: (i) suspension for
fine grains (smaller than 100$\mu m$), (ii) saltation and (iii) reptation
for grains of intermediate size (between 100 and 200 $\mu m$).}
\label{saltation}
\end{figure}

An important preliminary question is in order. Indeed,
even without  evoking the possibility
of writing basic sand equations,  we may ask a fundamental
question 
about locality-versus-nonlocality
in the aeolian sand ripple formation. More precisely does
dynamics of a given region (small in comparison to the ripple wavelength) depend
on that of a distant region located at a distance which is significantly
larger than the ripple wavelength? If so we can say that sand surface
dynamics must be nonlocal. %, otherwise dynamics would  local.
This question is still  controversial, and it seems
to us very important to settle it up from the very beginning.
The argument in favour of nonlocality rests
on the following fact:
because the grains that make a high fly (the saltation grains -- 
Fig.~\ref{saltation})
possess a saltation length $l$ which is much larger than
the ripple wavelength $\lambda$,
then a rather distant region on the sand bed would
get these saltating  particles. As  information is being passed
from two quite distant points, we would a priori think
of the importance of nonlocality. In reality, 
%for sand ripples there are two  kinds of grains:
the moving grains in the ripple formation process can divided into
two main categories:
the saltating ones that have a  high kinetic
energy (and make long jumps)
and the low energy splashed grains (dislodged by the impact 
of saltating grains -- see Fig.1)
which in turn travel in a hopping %(short hopping)
manner on a scale $a$ which is several (typically 6--10)
times smaller that the ripple wavelength. 
If nonlocality is adopted we must answer these two experimental facts
which clearly contradict it:
(i) as noted by Bagnold the saltating grains population
arrives on the sand bed almost at the same angle everywhere along the bed
-- as if a rain of particles were sent from a very long altitude
at a  fixed incident angle
(which is of order of $10^\circ$); so when they impact
on a region there is no way to distinguish between, say,
two gains that originate from two different regions
of the sand surface.
(ii) Additionally, as a  grain has been extracted
from the bed, becoming thereby a saltating one, it is transported by a turbulent
flow where at a such high Reynolds number the coherence length is so small that
during the fly the grains loose, so to speak, the memory of where it comes from.
Given these two facts it is hard to believe that saltating grains provides
any effective interaction between the topography of   two distant regions
on the surface. Thus
it seems  difficult to be in favour of nonlocality, albeit saltating grains
make, beyond any doubt, long jumps. Their high jumps simply imply that
their energy is such that they can dislodge some grains (the reptating grains)
and make them jump on a length scale $a$. If the saltating grains had
a higher energy, then $a$ would be increased which in turn (as seen also
experimentally~\cite{Anderson87}) would increase the ripple wavelength, keeping $\lambda/a$
to a typical value of order $6-10$.
In other words, and 
as again noted by Bagnold\cite{Bagnold41}, and reported by Anderson et al.\cite{Anderson87},
the saltating grains serve merely to bring energy into the system, 
the saltating 
population  exchanges almost no grains with the reptating population.
The ripple formation depends basically on the local topography of surface,
and the information is propagated only by reptating grains.
We believe that
we can even conceive the following experiment: we use an air gun %which is
designed to throw beads and 
inclined with some angle with  respect to an intially flat bead surface.
Then we move the air gun over the bed in a erratic fashion,
back and forth,
and send beads on the bed.
After a sufficiently large number of collisions 
the bead surface should develop a ripple structure.
By this way we completely
eliminate any notion of saltation; the air gun is simply injecting energy
into the system.

The  initiation of sand ripples as imagined
by Bagnold is appealing (see later).
However, a question of major importance has remained open
until recently: once the instability takes place, what is the subsequent
evolution of the instability? Would the instability lead
to an ordered or disordered structure?
Would the wavelength be that corresponding to
the linearly fastest growing mode?
The initiation of the instability is based on a linear analysis,
while the subsequent behaviour requires a non-linear treatment.
In the absence of any continuum theory of sand,
we have recently briefly discussed the derivation of a non-linear evolution
equation by evoking conservation laws and symmetry arguments
\cite{Csahok99a}. If the local
character is admitted, on the basis of many obvious
experimental facts described above, 
our equation should generically  be of the form  given in~\cite{Csahok99a}.
Any more or less microscopic theory of sand should, in the continuum limit,
be compatible with that equation.  It has been shown indeed that
using a hydrodynamical model for sand flow~\cite{Valance99} the derived
equation is that inferred from symmetry and conservation.
The aim of this paper is three-folds. (i) We give an extensive discussion
on the derivation of the non-linear evolution equation 
both from symmetry and conservation.
We shall also revisit the hydrodynamical
model and show that altering the basic model leads to a modified
equation precisely as dictated by symmetry and conservation.
(ii) We analyse in details 
%the far reaching consequences of the
the properties of the
continuum equation. It will be shown that this equation leads
to coarsening -- at the initial stage the linearly unstable mode
prevails, while at a subsequent times the structure coarsens. We shall
analyse the coarsening process and quantify the  exponent for the wavelength
increase in the course of time.
This task will be dealt with both analytically and numerically. We shall
discuss some variants of the originally proposed equation, and the contribution
of higher order terms. Though the quantitative feature may change, we shall
see  that the overall qualitative behaviour remains the same.
(iii) Since
there is a sparse information on sand ripples, and that various models
have been suggested in the literature, we have felt it worthwhile
to devote a  review to previous works,  and first to summarize
the basic  features and order of magnitude  of the underlying physical
phenomena of interest.
Thus the first part of the paper must be regarded rather
as a short review paper. 

This paper is organized as follows. In Section 2 we outline the main
physical ingredients in the formation of aeolian sand ripples.
Section 3 is devoted to a short review of different models.
Section 4 reconsiders  the hydrodynamical model from which
we can extract a non-linear evolution equation. We discuss
in particular different variants and its impact on the form
of the evolution equation. Section 5 uses
symmetry and conservation arguments to write down a generic evolution
equation and its variants. We shall then pay a special attention
to the analysis of the equation and its far reaching consequences.
Section 6 contains a summary and discussion.

\section{Outlines of aeolian sand transport}

According to Bagnold's vision\cite{Bagnold41}, 
the aeolian sand transport can be described in
terms of a cloud of grains leaping 
along the sand surface, the grains regaining from the wind the energy lost
when rebounding. His perception of the process, although refinements
and modifications have been developed in recent years
(see \cite{Anderson91} for a review of recent progress), still holds
in the main lines.

When the wind blowing over a stationary sand bed becomes sufficiently strong,
some particularly exposed grains are set in motion. Some grains
are lifted  by the pressure difference between 
the top and the bottom.
Once lifted free of the bed, the grains  are much more easily 
accelerated by the wind. Therefore as they return to the bed, some of the 
grains will have gained enough energy so that on impact they rebound 
and eject other grains. 

{\it Saltation} is usually defined
as the transport mode of a grain capable of splashing up
other grains. One can think of the saltating grains as the high-energy
population of grains in motion.
In the initial period after the wind set in,
the number of ejected 
grains resulting from one impact is on average larger than one. 
These ejected grains are generally
sufficiently energetic to enter in saltation. Therefore
the number of saltating grains increases at an exponential rate. As the
transport rate increases, the vertical wind profile 
is modified due to the presence of the curtain of saltating grains. The
wind speed drops so that the saltating grains are accelerated less and
impact at lower speeds. As a result, the number of grains ejected
per impact decreases and when it falls to one, an equilibrium is reached.
This equilibrium state is only stationary in a statistical sense. The number
of saltating grains may fluctuate around the equilibrium value.
One should note  however that it is possible that in the equilibrium state 
a few grains  are still dislodged into
saltation by fluid lift. The number of grains ejected per impact
would then be slightly smaller than one.

%It is possible that the mean replacement capacity is lessened 
%because the characteristic of the impact process are modified  by the
%increasing bombardment of the sand bed and by the high concentration
%of low-energy ejected grains in the air close to the surface. 

The cloud of grains transported in equilibrium state does not
consist of saltating grains only. On impact the saltating grains splash
up a number of grains  most of which do not saltate, i.e., their
energy is so low that, as they return to the bed, 
they can not rebound or eject 
other grains. The motion of these low-energy ejectas is usually 
called {\it reptation}.

In summary, in equilibrium state of transport, two populations of grains
can be distinguished: (i) the high-energy saltating  grains which  
travel by successive jumps over long distances and (ii)
the low-energy reptating grains generated upon impacts of saltating grains
which move over much shorter distances. 

The purpose of this section is to present an overview of the
current knowledge of aeolian transport.
%the physical mechanisms 
%responsible for the grain motion. 
We will first recall the characteristics of the wind
profile over a  flat sand surface and report the
modification induced by the presence of a saltation cloud. 
Then we will present the mechanisms of the initiation
of sand motion.  Finally, we will expose the main features
of the saltation and reptation motion.

\subsection{Wind profile}

When a flow of air is blowing over a flat rough surface,
the wind profile is defined by the standard form for
a turbulent boundary layer
\footnote{ One can recover the expression of the velocity field of
a turbulent flow near a wall by using simple physical arguments.
In the region near the wall, the flow is completely characterized
by the three following parameters: the shear velocity $U^*$, the
distance from the wall $z$, and the kinematic viscosity $\nu$.
However, the viscosity is important only very close to the wall.
One can therefore say that the mean velocity gradient
depends only on $U^*$ and $z$. Using dimensional analysis, one
obtains finally  the expected result.}
\cite{Tritton}
\beq
\frac{du(z)}{dz}=\frac{U^*}{kz} \kom
\eeq
where $u(z)$ is the average horizontal component of the velocity, 
$k$ is the Karman constant and 
$U^*$ is related to the shear stress on the ground 
\beq
\tau=\rho_a U^{*2} \kom
\eeq
$\rho_a$ being the density of air.
The wind profile can thus be written as 
\beq
u(z)=u_0 +\frac{U^*}{k} \ln \frac{z}{z_0} \kom
\label{windpro}
\eeq
where $u_0$ is the velocity at the reference height $z_0$ which
can be chosen at  convenience.
%$z_0$ is defined as the effective roughness of the bed surface. 
%($u_0$ is the velocity at height $z_0$). 
In absence of saltation cloud, $z_0$ is often chosen to be the 
roughness height of the bed surface.
A good estimation of $z_0$ for a flat sand surface is 
given by $z_0\simeq d/30$, where $d$ is the grain diameter\cite{Bagnold41}. 
At this height, the wind velocity is zero (i.e., $u_0=0$).
In a log-log plot, the height-velocity lines (associated to 
different wind strengths) all converge to
a focal point located at  the roughness height $z_0$ where the velocity
is zero.
The presence of saltation alters significantly the nature of the velocity
profile. For saltating flows, as experimentally shown by 
Bagnold \cite{Bagnold41}, 
the height-velocity lines
are also straight but converge at a different focus at some
greater height $z_0^{\prime}$ ($\simeq 5d$) and 
non-zero velocity $u_0^{\prime}$.

%However, the exact modification of the wind profile by the presence
%of saltation is still matter of debate. 
Concerning the incidence  
of a wavy sand bed on the wind profile, only little is 
known. The literature is quite poor on this topic. 
It should however be
interesting to have reliable data about the modification of the wind
profile by the presence of a ripple field.

\subsection{Initiation of sand transport}

The initiation process requires grains to be entrained by wind forces.
This occurs when the wind strength rises to the so-called
threshold fluid velocity to be defined below.

The initiation of grain motion can be understood by examining
the forces acting on individual grains.
Wind blowing over a sand surface
exerts two types of forces. (i) a drag
force acting horizontally in the direction of the flow,
and (ii) a lift force acting vertically upwards. (iii) Opposing
these aerodynamic forces are inertial forces, the most important
is  the grain's weight.

\begin{itemize}
\item The drag force is composed by
the friction drag and the pressure drag. The latter results 
from increased pressure on the upwind
face of the grain and decreased pressure on its downwind side. 
The friction drag is the viscous stress acting tangentially
to the grain. The total
drag
\footnote{It is
worth noting that the expression of the drag force together
with the lift one discussed below can be
established by means of a simple dimensional analysis. Indeed, if one
wants to write the drag force on a grain of size $d$ taking into
account the fluid velocity $U$, its viscosity $\nu$ and its density $\rho_a$,
the only way is  $F= f(R)\rho_a d^2 U^2$ where $f$ is a function of
the Reynolds number $R=Ud/\nu$. At high Reynolds
number (that is the case we are dealing with), 
the  force is expected to be independent of the fluid
viscosity (at the scale of the grain the effective
Reynolds number is too large) so that $f$ must be independent of $R$. On the contrary at
low Reynolds number, the  force is viscosity-dependent and inertia
must scale out of the equation. So the  only way is that $f$ must scale
as $1/R$. We thus recover the Stokes law.}
acting on the grain is given by
\beq
F_d =  \beta \rho_a d^2 U^{*2} \pnt
\eeq
$d$ is the grain diameter and
$\beta$ is a parameter depending on the Reynolds number $R^*=dU^*/\nu$.

\item The lift force (or Magnus-Robbins force)
is inherent to Bernoulli effect. Indeed, it arises
because of the high wind velocity gradient near the bed. The flow
velocity on the underside of a grain at rest  on the bed is zero but
on the upper side the flow velocity is positive. Due
to Bernoulli's law this leads to an underpressure
on top of the grain, causing a lift. The average lift 
force can be expressed as
\beq
F_l= C_l \rho_a d^2 u^2 \kom
\eeq
where $u$ is the fluid velocity evaluated at the top of the grain and
$C_l$ is a lift coefficient which depends
on the Reynolds number $R=du/\nu$. Note that $u$ can be easily related
to the shear velocity $U^*$ thanks to eq.~\ref{windpro}.

\item Finally, the effective weight of a grain immersed in a fluid is
given by
\beq
P=\rho_g^{\prime} g d^3 \kom
\eeq
with $\rho_g^{\prime}=\rho_g-\rho_a$, $\rho_g$ being the density of the grain.
The fact that $\rho_g-\rho_a$, enters the weight  and not $\rho_g$
is due to the Archimedes force.
\end{itemize}

One can now examine the balance between the different forces acting on
individual grains. Consider a flat surface covered by loose sand of uniform
size. Grains in the top layer of the bed are free to move upward but
their horizontal movement is constrained by adjacent grains. The point
of contact between neighbouring acts as a pivot around which rotational
movement takes place when the lift and drag forces exceed the inertial
force. The threshold at which the grains detach from the ground is then
reached when the moment of the three forces about the pivot 
balance each other
\beq
(d/2)(F_l+F_d)=(d/2)P
\eeq
This corresponds to a threshold shear velocity $U^*_t$ determined by
\beq
U^*_t = A^* \sqrt{\frac{\rho_g -\rho_a}{\rho_a}gd}
\eeq
where $A^*$ is a coefficient which depends essentially 
on the Reynolds number $R^*$. 
$A^*$ turns out to be fairly constant when the Reynolds number $R^*$ is
large compared to $1$. 
The threshold value of $U^*$ for fine
dune sand with a diameter of $0.02$ cm is about $0.2$ m/s 
and the corresponding
value of $R^*$ is of order of unity (see \cite{Bagnold41} for
more details). 
For this and for all sands of larger
grain size, $A^*$ is found to be  constant  (in air $A\simeq 0.1$).

\subsection{Saltation motion}

The first stage in the grain motion is lifted, discussed
before. After being  lifted, grains are transported by wind 
and start to make successive long jumps (that is saltation motion). 
Then an equilibrium establishes between
the saltating grains and the wind profile.
According to Bagnold, the motion of the saltating grains can
be described by an average trajectory. 
In particular, one can define an average height and length of the
trajectory of the saltating grains. 
These quantities can be estimated considering
that the moving grain experiences the gravity force 
$\rho_g^{\prime} d^3 \mathbf{g}$,
the air friction $C_d \rho_a d^2 V_r \mathbf{V}_r$ 
(where $V_r$ is the relative
velocity of the grain with respect to the air) and the lift force
$C_l \rho_a d^2 V_r^2 \mathbf{n}$ (where $\mathbf{n}$ is a unit vector
perpendicular to $\mathbf{V}_r$).
The equation of motion of a grain therefore reads , after projection
on the horizontal and vertical axis, 
\beqa
&& \rho_g d^3 \frac{du_p}{dt}=  - C_d \rho_a d^2 V_r(u-u_p)
+C_l \rho_a d^2 V_r v_p \kom \\
&& \rho_g d^3\frac{dv_p}{dt}=-\rho_g^{\prime}d^3 g- C_d \rho_a d^2 V_r v_p
-C_l \rho_a d^2 V_r(u-u_p) \kom
\eeqa
where $u$ is the horizontal velocity of the air,
$u_p$ and $v_p$ are the horizontal and 
vertical components of the grain velocity. 
The relative velocity is given by $V_r=\sqrt{(u-u_p)^2+v_p^2}$.
These are
two coupled non-linear equations for which
an analytic solution does not look possible. However, making
some simplifications, it is possible to extract 
the basic features of the grain trajectory. 
We will therefore assume that (i) the lift
force is negligible \footnote{ The lift force
is expected  to play a role only in the region near the bed.},
(ii) the wind velocity is uniform
along the vertical axis $z$ and equal to 
 $\bar{u}$, and 
(iii) the horizontal particle velocity
is rapidly in equilibrium with the wind flow (i.e, $u_p \simeq \bar{u}$).

Making use of these assumptions, it is straightforward
to calculate the height $h_s$, 
the length $l$ and the incidence angle $\alpha$ of the saltating grain
trajectory:
\beqa
h_s&=&\frac{v_{\infty}^2}{2g} \ln (1+\frac{v_0^2}{v_{\infty}^2}) \kom \\
l&=&\bar{u}t_f \kom \\
\tan \alpha&=& 
\frac{v_{\infty}}{\bar{u}\sqrt{1+\frac{v_{\infty}^2}{v_{0}^2}}} \kom 
\eeqa
where $t_f$ is the time of flight
\beq
t_f=\frac{v_{\infty}}{g} \left[ \arctan (\frac{v_0}{v_{\infty}})
+ \ln (\sqrt{1+\frac{v_0^2}{v_{\infty}^2}}+\frac{v_0}{v_{\infty}}) \right]
\pnt 
\eeq
Note that $v_{0}$ is the liftoff velocity 
and $v_{\infty}$, the terminal velocity of a grain falling in air, is
given by 
$v_{\infty}=\sqrt{\rho_g dg/C_d \rho_a } $.
For typical fine sand grains with a diameter of $0.02$ cm, 
the terminal velocity is about $1$ m/s. One should point out
at this stage that the liftoff velocity $v_0$ 
corresponds to the speed of saltating grains
just after their rebound on the granular bed. In the equilibrium
state of saltation, this velocity
is quite large (of order of a few $m/s$) since the saltating grains
have been able to extract energy from the wind during their flight.

In the limit where $v_0$ is appreciably larger  than $v_{\infty}$, we
get the following simple results:
\beqa
h_s&=& \frac{v_0^2}{2g} \frac{\ln(v_0^2/v_{\infty}^2)}{v_0^2/v_{\infty}^2}
\kom \\
l&=& \frac{2 \bar{u}v_0}{g} \frac{\ln (2v_0/v_{\infty})}{2v_0/v_{\infty}}
\kom \\
\tan \alpha &=& \frac{ v_{\infty}}{\bar{u}} \pnt
\eeqa
If we take $v_0=2.5$ m/s, and $\bar{u}=5$ m/s,
we find $h_s\simeq 10$ cm, $l\simeq 80$ cm and $\alpha \simeq 12^{\circ}$
for grain size of $0.02$ cm.
The following remarks are in order. (i) One can note
that the values of the hop  length and height are less than those
expected in absence of vertical drag. (ii) The hop height relative
to $v_0^2/2g$
decreases with the liftoff velocity. (iii)
The hop length increases both with the wind strength and the liftoff
velocity. (iv) Finally, the impact angle is independent
of the liftoff  velocity. These results, although
derived in a crude way, give the same trends as those found
from the full numerical calculation.

One should point out that we have omitted here
the Magnus lift force which is present
where grains are rotating.  This additional force
can enhance significantly the height and length of the saltating 
trajectory \cite{White77}.
However, the general features found above remains
qualitatively correct (see for example \cite{Anderson86}).

\subsection{Collision process}

The collision between the saltating grains and the bed surface
is a crucial process in the aeolian sand transport. 
Although aeolian transport is
initiated by aerodynamic forces as seen above, its maintenance
relies essentially through impacts. In other words,
the dislogment of grains from sand bed is essentially induced
by the impacts  of saltating grains.
The collision process between the saltating grains
and the sand bed is of great importance  both in saltation 
and reptation motion. In particular, the rebound angle and
the liftoff velocity of the saltating grain, as well as the
dynamics properties of the ejected grains (i.e., reptating grains)
can only be determined by a thorough knowledge of the collision process.

Recent experiments \cite{Mitha86,Willets86}
and numerical simulations \cite{Ungar87,Rumpel85,Werner88,Werner90}
have focused on the saltation impacts. It is worth recalling
the main results.
The impact of a single saltating grain typically results in
one energetic rebounding grain and a large number of emergent grains (i.e.,
low-energy ejectas or reptating grains). The rebounding
grain leaves the surface with roughly two-third of the impact velocity
while the emergent grains have a mean ejection speed less than
$10\%$ of the impact speed and therefore have a short
trajectory in the air. Change of impact angle and 
speed is found to affect the outcomes
of collision in several ways. 

First,  with decreasing angle of incidence
the ratio of the rebound to impact vertical speed increases. This
ratio is greater than one for low incident angles ($\sim 10^{\circ}$) which
correspond precisely to impact angles observed in the equilibrium state
of saltation transport. 
It  means that the saltating grain can reach a height as great as that 
from where it fell
provided that  the amplification of the vertical
speed is sufficient to overcome drag losses as it rises. 
If this condition is achieved, the saltation is able
to maintain itself due to the particle impact.  

Second,
the properties
of the emergent grains (speed and take-off angle) are practically
unaffected by a change of the incident angle or impact speed. 
The only noticeable effect is that an increase of the impact speed
results in an augmentation of the number of the ejectas.

Although the understanding of the collision process has been
largely improved in the last decades, there remains an important
set of problems to deal with. (i) Both theoretical and experimental
works assume that the bed is comprised of a single grain size
which is obviously not the case in the real word.
(ii) In most of studies, the sand bed is taken as simple as possible:
flat and uniformly packed. However, it seems clear that 
the local bed topography (at the ripple scale) 
as well as the variation of the bed packing fraction may alter significantly
the nature of the collision process. 
(iii) Finally, there is an important limitation in almost all
studies: the third dimension is ignored. A three-dimensional knowledge of
the collision process would however be necessary for the understanding of the
complex spatial evolution of a 3D bed over which saltation occurs.

\section{Short review of analytical models }

Once we have clarified and summarized the main physical
aspects and order of magnitudes in the process by which 
ripples form, we are in a position to tackle the  problem
of ripples itself. 
We present here a survey of analytical models of
aeolian ripple formation  proposed in the literature.
We shall also suggest how these models could be modified
in order to include a more realistic dynamics both in the linear
and non-linear regimes. This step of the extension of the model will serve
as a preliminary analysis before tackling the full non-linear analysis
in the subsequent sections, which is the main topic.

All these models explain the ripple formation as the result
of a dynamical instability of a flat sand bed.
However, one should point out that two fundamentally
different explanations for the
ripple instability have been proposed in the literature.
The first one is based on the fact that
the reptation flux varies according to the local slope of the bed profile
and has been developed by Bagnold\cite{Bagnold41} and later on by 
Anderson\cite{Anderson87}. 
Another explanation  has been proposed by Nishimori and Ouchi\cite{Nishimori93b}
based on the variation of the saltation length according
to the height of the sand bed. This second theory, although it is
appealing, is not really supported by experimental evidences.

\subsection{Anderson Model}

As it will emerge the key ingredient of the flat bed instability 
is of geometrical nature: an inclined surface is subjected to
more abundant collisions than a flat one. That is to say the
 the mass current is an increasing function of the slope. 
This is a  situation  where the  consequence acts in favour of the cause,
 say in a way which is against the Lechatellier principle, leading 
 thus inevitably to an instability, as seen below.

In the Anderson model\cite{Anderson87}
the saltating grains are not directly responsible for the ripple
instability. 
Instead, the saltating grains
are just considered  as an external reservoir
which brings energy into the system. 
They are assumed to be sufficiently
energetic that they can travel on large
distances without being incorporated into the sand bed. Furthermore,
at each impact, they can eject  
a certain number of low-energy grains (i.e., reptating grains)
which make single small hops. The ripple instability
is driven by the the flux of the reptating grains.

The surface is assumed to be subject to a homogeneous
rain of saltating grains impacting with a uniform incident angle.
On impacts the saltating grains eject low-energy grains which hop 
over a characteristic reptation
length $a$. The local sand height $h(x,t)$ changes
in the course of time, this is simply due to the fact
that a horizontal flux of particles exists. Due
to mass conservation  we must have
\beq
\frac{\partial h}{\partial t}=-\frac{1}{\rho_g}\frac{\partial Q}{\partial x}
\label{eq0} \kom 
\eeq
where $Q$ is the horizontal mass flux of moving grains 
(i.e., the mass of grains per unit time and unit width of flow). 
The horizontal flux can be split into
two contributions (i.e, the flux of saltating grains and reptating
grains):
\beq
Q(x)=Q_s + Q_{rep}(x) \kom
\eeq
with
\beq
Q_{rep}(x) = m_p  \int_{x-a}^{x} N_{ej}(x) dx \kom
\eeq
where $m_p$ is the mass of a grain, 
and $N_{ej}(x)$  the number of ejected grains
at $x$ per unit time and surface to be specified below. 
Taking advantage of the expression of the flux, the governing equation 
for the bed profile [eq.~(\ref{eq0})] can be rewritten as
\beq
\partial_t h=-d^3 [N_{ej}(x)-N_{ej}(x-a)] \pnt
\label{eq1}
\eeq
We recall that $d$ is the grain size.

\begin{figure}
\begin{center}
\includegraphics[width=12cm]{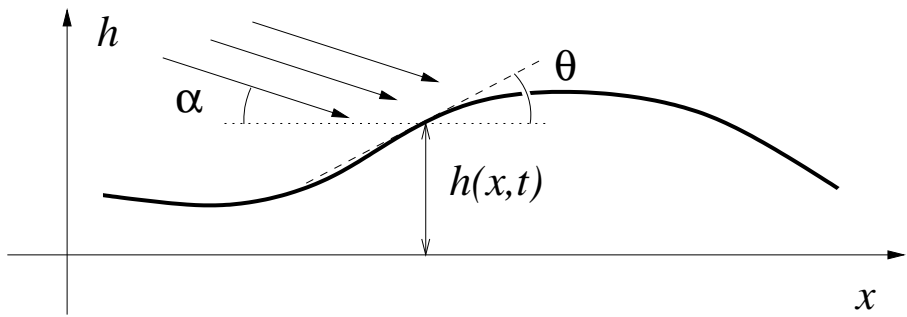}
\end{center}
\caption{Example front configuration.}
\label{explain}
\end{figure}

The rate of ejected grains can 
 directly be related to the number $N_{imp}$ of impacting
(i.e., saltating) grains: 
\beq
N_{ej}(x)=n_0  N_{imp}(x) \kom 
\label{nej}
\eeq
where
$n_0$ is the number of grains ejected per impact. 
In a first approach, $n_0$ can be taken as 
constant whereas $N_{imp}(x)$ clearly depends  
on the impact angle of the saltating grains with respect
to the local bed slope at the position $x$.
If $\alpha$ measures the impact angle of the saltating grains
with respect to the horizontal and
$\theta$ the angle of the local bed slope (see Fig.~\ref{explain}), the rate of
impacting grains reads
\beqa
N_{imp}&=& N_0 \cos \theta (1 +\frac{\tan \theta}{\tan \alpha})
\nonumber\\
&=& N_0 \frac{(1 +\cot \alpha\; \partial_x h)}{ [1+(\partial_x h)^2]^{1/2}}
\label{Nimp} \pnt 
\eeqa
$N_0$ is the number of saltating grains arriving on a flat horizontal
bed per unit time and unit  surface.

\begin{figure}[h]
\begin{center}
\includegraphics{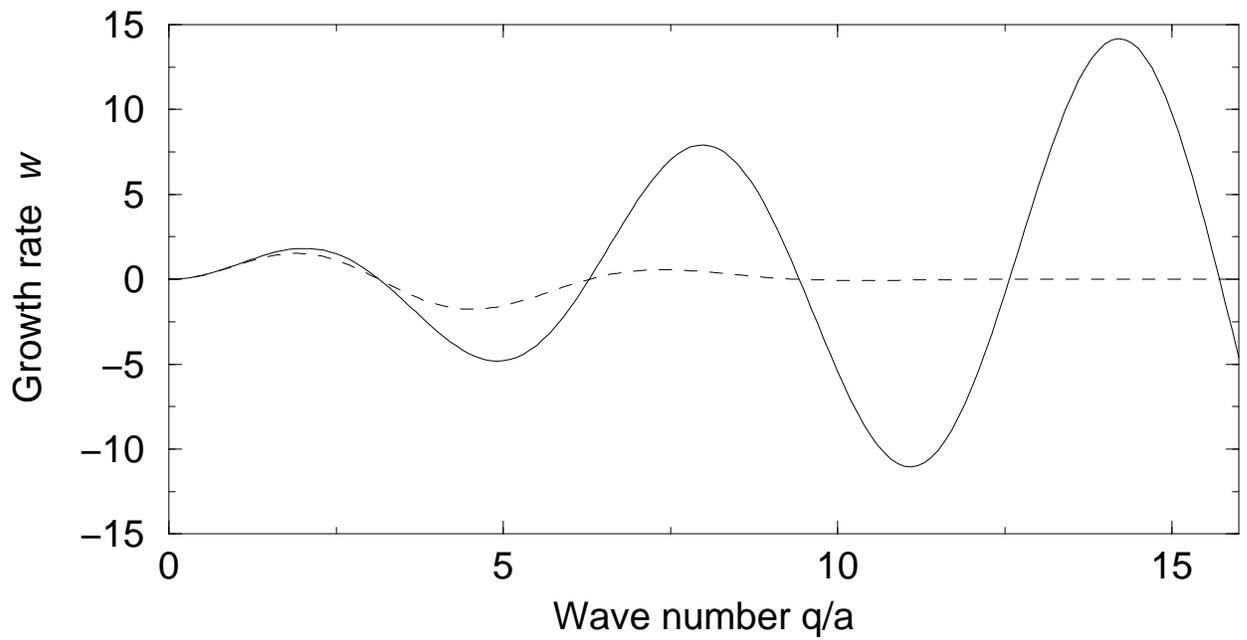}
\end{center}
\caption{
Dispersion relation in the framework of Anderson's model.
Real part of the growth rate as a function of
the wave number. Full line: spectrum in the case where the
reptation length is constant.
Dashed line: spectrum in the case where the reptation length is
distributed according to a Gaussian law.
}
\label{andersp}
\end{figure}

Eq.~(\ref{eq1}) together with
(\ref{nej}) and (\ref{Nimp}) completely describe the evolution
of a sand bed surface subject to saltation. A flat profile
is obviously solution of this equation but the question of
interest is to know whether it is stable
against small fluctuations. 
First we extract the leading term in $h$ from $N_{imp}$ and inject it
into the equation for $h$.
Then seeking solutions 
of the form $h\sim e^{iqx+\omega t}$ (where $q$ is the wave number), we obtain
\beq
\omega=-\mu_0 \cot \alpha\; i q[1-e^{-iqa}] \label{spectre1}
\kom 
\eeq
where $\mu_0=n_0 N_0 d^3$.
One can see (cf. Fig.~\ref{andersp}) that there  is
an infinite number of bands of unstable modes. 
The flat bed surface is unstable. One can also note
that each band exhibits a maximum at $k=(4n+1)\pi/2a$
and the growth rate of these maxima diverges for large wavenumber
which is physically not acceptable.

Anderson refined his model to circumvent this problem
by introducing a dispersion in the reptation length. 
If we call $p(a)da$ the probability
that the reptation length is comprised between $a$ and $a+da$,
the governing equation~(\ref{eq1}) for the bed surface becomes
\beq
\partial_t h= -d^3 \int_{-\infty}^{\infty} p(a)[N_{ej}(x)-N_{ej}(x-a)]da
\pnt
\eeq
In that case, the linear stability analysis yields
\beq
\omega=-\mu_0 \cot \alpha \; iq [1-\hat{p}(q)]
\kom
\eeq
where $\hat{p}(q)$ is the Fourier transform of $p$.
If we assume that $p(a)\sim e^{-(x-\bar{a})^2/2\sigma^2}$ 
(where $\bar{a}$ is the mean reptation length and $\sigma$ is the
variance), we get
\beq
\omega= -\mu_0 \cot \alpha\; iq \left[ 1-e^{-iq\bar{a}}e^{-\sigma^2q^2/2}\right]
\eeq
The flat surface is again unstable (see Fig~\ref{andersp})
but contrary to the previous case, the most dangerous mode
has a finite wavenumber. In the case where
the dispersion is large enough (i.e., $\sigma > \bar{a}$),
the most dangerous mode is the first peak at $q=\pi/2\bar{a}$ which corresponds
to a  wavelength $\lambda=4\bar{a}$. This mode is expected
to dominate the subsequent development of the instability and therefore
to give the order of magnitude of the ripple wavelength.
One can conclude that the
dispersion in the reptation length damps
the growth of the large wavenumber modes.

One shall add a few comments about the Anderson model. It can
be interesting to rewrite Anderson equations in the limit
where the reptation length is smaller than the bed deformation
(i.e., the ripple wavelength). This limit corresponds to the
usual situation encountered in the case of aeolian sand ripples. 
In other words, the process of ripple formation can be considered
as a local one.
In this limit, one can perform a Taylor expansion of
$N_{ej}(x-a)$ about the position $x$ and the governing 
equation~ (\ref{eq1}) can be approximated by
\beq
\partial_t h \simeq  -d^3 \left[ a \partial_x -\frac{a^2}{2} \partial_x^2
+ \frac{a^3}{6} \partial_x^3  + \dots \right]  N_{ej}(x) \label{eq2} \pnt
\eeq
Using the expression of $N_{ej}$ and keeping only the linear
terms, one gets
\beq
\partial_t h = f_1 \partial_{xx} h + f_2 \partial_{xxx} h + f_3 
\partial_{xxxx} h  \kom
\label{eq3}
\eeq
with $f_1=-a\mu_0 \cot \alpha$, $f_2=(a^2/2) \mu_0 \cot \alpha$, 
and $f_3=-(a^3/6) \mu_0 \cot \alpha$. 
Note that the first linear term of the right hand side (whose coefficient
is negative) is directly responsible for the ripple instability.
The third derivative term is inferred to the drift of the ripple 
structure whereas the last one stabilizes structure at large
wavenumber. The growth rate of a mode $q$ is given
by  $\omega= a \mu_0 \cot \alpha [q^2 -i(a/2)q^3-(a^2/6) q^4]$
which is nothing but the long wavelength limit of expression~(\ref{spectre1}). 
The wavelength of the most dangerous mode is here of the same order of that
found previously: $\lambda=2\pi a/\sqrt{3} \simeq 4a$.

The Anderson Model gives a good description for the initiation
of the ripple instability but it is not intended to
predict the subsequent non-linear dynamics of the sand  bed profile, which
we shall attempt to consider here. 

\subsection{Hoyle-Woods Model}

The Hoyle-Woods  model \cite{Hoyle97} is an extension  of the Anderson model. 
Hoyle and Woods 
have taken into account the rolling and avalanching effect of the 
grains under the influence of the gravity as well as the shadowing
effect. However we shall forget here avalanching which is generally
absent in the process of aeolian ripple formation (slip faces are solely
observed on the lee slope of dunes).

The rolling effect can be important on the lee slope of the ripple.
Indeed, the reptating grains can roll down a slope under the influence
of gravity. This can be modeled by an additional
horizontal flux $Q_{rol}$
\beq
Q_{rol} = m_p N_r u_r \cos \theta \pnt
\eeq
$N_r$ is the number of rolling grains per unit surface and
$u_r$ is the speed of the rolling grains along the slope. 
The authors assumed that $N_r$ is constant and $u_r$ is
a function of the gravitational force
\beq
u_r= -\frac{\sqrt{gd}}{r} \sin \theta \kom
\eeq
where $r$ is a function of the grain packing and grain size.
Taking into account
of this additional flux, the governing equation for the bed height
is given by
\beq
\frac{\partial h}{\partial t}=-\frac{1}{\rho_g}
\left[ \frac{\partial Q_{rep}}{\partial x}
+ \frac{\partial Q_{rol}}{\partial x} \right]
\pnt 
\eeq
In the long wavelength limit (where the wavelength of the ripple structure
is much larger than the reptation length) the
bed growth due to reptation motion can be approximated by
$[\partial h/\partial t ]_{rep} \sim \partial_x N_{ej}$,
retaining only the leading order term (see eq.~\ref{eq2}).
In this limit, the evolution equation 
takes thus the following form
\beq
\frac{\partial h}{\partial t}= -\partial_x
\left[ a \mu_0 
\frac{(1+ \cot \alpha \, \partial_x h)}{[{1+(\partial_x h)^2}]^{1/2}}
- \nu_0 \; \frac{\partial_x h}{[1+(\partial_x h)^2]} \right] \kom
\eeq
where $\mu_0= n_0 N_0 d^3$ and $\nu_0= N_r(\sqrt{gd}/r)d^3$. 
An expansion in power of $h$ yields to leading order
\beq
\frac{\partial h}{\partial t}= 
f_1 \partial_{xx} h
\kom
\label{eq5}
\eeq
with $f_1= (\nu_0 -\mu_0 \cot \alpha)a$.
One clearly sees that the rolling effect introduces
a threshold for the ripple instability. Indeed the
flat bed surface is unstable only
if $\mu_0 \cot \alpha > \nu_0$. 
We recall that 
$\mu_0$ represents the flow rate of  reptating grains  for a flat sand
surface ($\mu_0 \cot \alpha$ is nothing but the excess flow rate when
the sand bed is tilted) whereas 
$\nu_0$ corresponds to the flow rate of rolling grains for a tilted surface.
The instability results therefore from a competition between reptation
and rolling motion.
As $\nu_0$ is assumed to be constant in this model, it follows that
high saltation flux (i.e., high value of $\mu_0$) or
low impact angle (i.e, small $\alpha$) 
favour the destabilization of the bed surface. 
In summary, 
the rolling effect tends to smooth out surface irregularities and
therefore the ripple instability can occur only above a certain threshold.

In this model, the ripple structure resulting from the instability
has no characteristic length (contrary to the Anderson model)
since the most dangerous mode is not finite. 
To circumvent this problem, Hoyle and Woods have taken into account
the shadowing effect. On the lee slope of
the ripple,
they considered that there  is 
a region beyond the ripple crest which is shielded from
the saltation flux. This is called the shadow zone and no hopping
occurs there. In the shadow zone the ripple evolves solely owing to rolling.
The role of this  shadowing effect has been investigated numerically by
Hoyle and Woods. They found stable ripple structure
whose wavelength is governed by  the length of the shadow
zone, as expected from simple geometrical considerations. 

\subsection{Nishimori-Ouchi Model}
\begin{figure}[h]
\begin{center}
\includegraphics{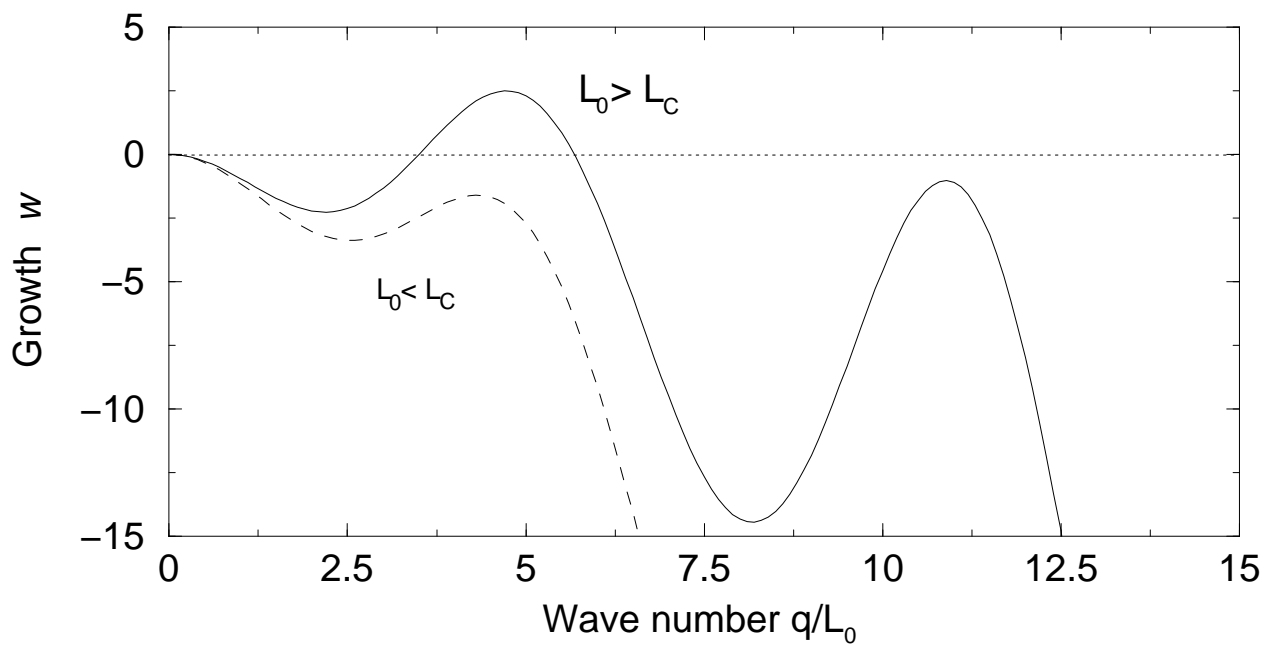}
\end{center}
\caption{
Relation of dispersion in the framework of Nishimori-Ouchi  model.
Real part of the growth rate as a function of the wave
number. Full line: above the instability threshold. Dashed line: below
the instability threshold.
}
\label{nishisp}
\end{figure}

The Nishimori-Ouchi model is based on the hypothesis
that the saltation flux is not homogeneous when the sand surface is 
deformed. They assume that the hopping length $l$ of the salting grains
depends on the location where they take off: 
\beq
l(x)=l_0+ b h(x) \kom
\eeq
where $x$ is the location of takeoff. Furthermore the saltating
grains are assumed to be incorporated to the sand surface when they
hit it. According to these hypothesis, 
the flux of saltating grains can be written here
as
\beq
Q_s(x)= m_p \int_{\zeta}^{x} N_s(x^{\prime}) dx^{\prime} \kom
\eeq
where $N_s(x^{\prime})$ is the number of saltating grains which
take off at  $x^{\prime}$
and $\zeta$ is the location of the takeoff point of the grains
which lands at $x$. The authors also consider  a reptation motion
(or creep) induced by gravity
\beq
Q_{rep}(x)=- \frac{D_r}{\rho_g} \frac{\partial h}{\partial x} \kom
\eeq
$D_r$ is a constant coefficient which stands for the relaxation rate.
The dynamics of the sand bed is thus given by 
\beqa
\frac{\partial h}{\partial t}&&= - \frac{1}{\rho_g}
\frac{\partial}{\partial x} (Q_s+Q_{rep})\\
                 &&=-d^3 (N_s(x)-\frac{\partial\zeta}{\partial x} N_s(\zeta))
			      +D_r \frac{\partial^2 h}{\partial x^2} \pnt
\eeqa
In the limit of small deformations of the bed surface,
one gets to leading order (assuming that $N_s=constant$)
\beq
\frac{\partial h}{\partial t} |_x= - b N_s \frac{\partial h}{\partial x}
|_{x-l_0} 
+ D_r \frac{\partial^2  h}{\partial x^2} |_x \pnt
\eeq
The growth rate of a  mode of wavenumber $q$ is then given by
\beq
\omega= -b d^3 N_s iq e^{-iql_0}-D_r q^2  \pnt
\eeq
One can easily show that the sand surface is unstable only if
$l_0$ is greater than a critical value given by $l_c=3 \pi D_r /2b d^3 N_s$
(see Figure~\ref{nishisp}).
Furthermore, near the instability
threshold the most dangerous mode $k_{max}\simeq 3 \pi/2l_0$
which corresponds to a wavelength $\lambda_{max}\simeq 4 l_0/3$.
In this model, the order of magnitude of ripple wavelength is  
given by the saltation length  whereas in the Anderson model
the pertinent length is the reptation one. Furthermore the
ripple wavelength is of the same magnitude of order as the saltation
length. The problem can not be therefore treated in the long wavelength
limit. Here the process of ripple formation can not be 
considered as a local one.  

The Nishimori-Ouchi model gives interesting results, however the way of
modelling the saltating grains can be seriously questioned.
First, the mechanism of ejection due to impacts of saltating
grains on the sand (whose importance has been evidenced by Bagnold 
\cite{Bagnold41})
is not taken into account, and second
the variation of the saltating length according to the takeoff point
has never been clearly set neither from wind tunnel experiments nor from
field observations. Furthermore, as we discussed in the introduction, 
it is hard to conciliate this picture with evidences  in favour
of locality.

\section{Hydrodynamic model}

We expose here a phenomenological model which is inspired
from the "BCRE" model\cite{Bouchaud94}
developed in the context of avalanches in granular flows.
This model has been adapted to the ripple formation process first by
Bouchaud and his coworkers \cite{Terzidis98} and later on
by Valance et al. \cite{Valance99}.
This model is based on a continuous description where
dynamics of the two pertinent grain species  (that is the moving
grains and the grains at rest) are considered. 
One of the advantage of the model
is to treat separately the erosion process and the deposition
one.  

This model has been  presented in \cite{Terzidis98,Valance99}.
However we find it worth recalling the main lines. This will  
allow us to discuss more critically the model and to
show how the final equation may be  sensitive to the starting
physical ingredients. This will clear up the question
of why the equation derived in  the next section (based
on symmetry) contains
additional nonlinearities not present in \cite{Valance99}, and
to show how this can be cured.

We shall call
the moving grains density $R(x,t)$ where $x$ is the coordinate
in the direction of the wind and $t$ the time. 
The grains at rest are measured in term of the local
height $h(x,t)$ of the static bed. In the thermodynamical limit, the
dynamical equations of $h$ and $R$ read
\beqa
&& \partial_t R= -V\partial_x (R)+ \Gamma[R,h] \label{eqR} \kom\\
&& \partial_t h= -\Gamma[R,h] \label{eqh} \kom
\eeqa
where $V$ the mean velocity of the moving grains and
$\Gamma$ describes the exchange rate between the moving grains and the 
grains at rest  
\beq
\Gamma=\Gamma_{dep}+\Gamma_{ej} \label{eqG} \kom
\eeq
the first term describing the deposition process of the reptating
grains and the other modelling the ejection of grains from
the bed surface. $\Gamma$ depends a priori on $h$ and $R$.

The expression of $\Gamma$ can be determined using
phenomenological physical arguments. 
We have seen that 
the saltating grains are never caught by the bed surface
but act as a reservoir of energy. They induce reptation motion
which is directly responsible for the ripple formation.
Among the moving grains, we are therefore interested only in those
in reptation.
$\Gamma$ should describe the exchange rate between the grains
at rest and the reptating grains. 

As seen above, the ejection rate of reptating grains is essentially driven
by the flux of the saltating grains hitting the surface.
To a smaller extent, one can expect that a small part of the
reptating population  are set in motion by the wind. One shall therefore
consider two ejection mechanisms, one due to impacts of the saltating
grains and the other driven by the wind.\\

\noindent(i) The ejection rate due to impacts  
can be modeled using Anderson approach where the
ejection rate is given by $\Gamma_{ej}^{imp} \sim n_0 N_{imp}$
(we recall that $N_{imp}$ the rate of saltating grains impacting 
the granular bed and  $n_0$ is the number of ejected grains per impact).
Anderson assumed $n_0$ to be constant. We will consider here
that the efficiency of the ejection can depend on the bed topography
and especially  on the bed curvature.
Indeed, it is natural to think of that
it is easier to dislodge a grain at the top of a bump than
in a trough. The number $n$ of ejected grains per impact 
can thus be modelled by
\beq
n=n_0(1-c\kappa) \kom
\eeq
where $c$ is a constant parameter and $\kappa$ the bed curvature.
The rate of ejection reads therefore
\beqa
\Gamma_{ej}^{imp} &&= d^3 n_0 (1-c \kappa) N_{imp}\\
&&= d^3 n_0 N_0 \left( 1-c \frac{h_{xx}}{(1+h_x^2)^{3/2}} \right)
\frac{(1+\cot \alpha \, h_x)}{(1+h_x^2)^{1/2}} \pnt
\label{Nej}
\eeqa
In the limit where $h_x \ll 1$, an expansion in power of $h_x$
can be performed. Retaining the terms up to the quadratic order
one gets
\beq
\Gamma^{imp}_{ej} =\alpha_0 (1+ \alpha_1 \partial_x h- \alpha_2 \partial_x^2 h)
- \alpha_0 
\left[ \alpha_3 h_x^2 + \alpha_4 \partial_x(h_x^2) \right] + O(h_x^3)
\label{ej1} \pnt
\eeq
$\alpha_0= d^3 n_0 N_0$, $\alpha_1=\cot \alpha$, $\alpha_2=c$,
$\alpha_3=1/2$ and $\alpha_4=c \cot \alpha$.
$\alpha_0$ is directly related to the number $N_0$ of saltating grains
hitting a flat surface per unit time and unit surface. 
Let us recall the meaning of the different terms.
The term proportional to $\partial_x h$
expresses the fact that the rate of ejection is greater when the local
slope is facing the wind (the flux of saltating grains being larger on
the stoss side as seen previously). 
The last linear term takes into account the curvature effect:
it is harder to dislodge grains in troughs than at the top
of a crest. There are  two nonlinear terms.
The first nonlinearity
comes from the contribution of the slope effect on the
ejection rate and the second one corresponds to the coupling between
slope and curvature effects. It is worth noting that these
nonlinear terms have been neglected in previous works \cite{Valance99} 
but turn out
to be important in the nonlinear development of the ripple instability
under some certain circumstances to be specified below.\\

\noindent (ii) The ejection rate due to wind entrainment is
in principle extremely weak because the wind is screened
by the saltating grains. Indeed, it has been found 
from numerical simulation that the fluid entrainment
is unimportant for a flat surface. However, one may think that
the direct dislodgement by the wind of a grain located
on the top of a crest can be significant. One
can thus assume that the ejection rate due to wind entrainment
is driven by curvature effect
\beq
\Gamma_{ej}^{wind}= -\beta_2 \partial_{xx} h + O(h_x^3)
\label{ej2}
\eeq

\noindent (iii)
The rate of deposition is assumed to be proportional to the number
$R$ of reptating grains so that we can write
\beqa
\Gamma_{dep}&&= - R \gamma \nonumber \\
&&=-R \gamma_0(1 \pm \gamma_1 \partial_x h +\gamma_2 \partial^2_x h)
\label{dep}
\eeqa
$\gamma^{-1}$ represents the typical time
during which the reptating grains are moving before being
incorporated to the sand bed. This life time can be interpreted
in terms of the characteristic reptation length $l$ which can be defined
by  $l=V/\gamma$, where $V$ is the mean speed of the reptating
grains. The first contribution in (\ref{dep}) represents the deposition rate
for a flat bed surface. As a consequence
$\gamma_0$ corresponds to the typical life time
of a reptating grain on a flat surface. The other contributions
mimic the slope and curvature effects. The effect of slope 
on deposition process can be different depending on  the importance
of the wind drag on the reptating grains. If one assumes that
the wind drag is negligible near the surface, one may
think
that the deposition process is enhanced on a stoss slope
(positive sign in front $\gamma_1$). Indeed, the reptation
length is expected to be smaller on a stoss slope due to gravity (cf.
Hoyle-Woods model).
On the other hand, if the wind drag near the bed surface is
significant, the deposition process should be weakened on slope
facing the wind (negative sign in front $\gamma_1$)
since a reptating grain on a stoss slope can
gain additional energy from the wind and therefore  travel
over a longer distance.
Finally, the plus sign in front
the term modelling the curvature effect clearly indicates that
the deposition is favoured in troughs in comparison to crests.

The set of equations~\ref{eqR},~\ref{eqh} and~\ref{eqG} plus
\ref{ej1}, \ref{ej2}, \ref{dep}
describe completely our system. There exists a trivial solution
corresponding to the situation where the bed surface is flat. In this
case, the density of reptating grains is simply given by
$R_0=\alpha_0/\gamma_0$. The next step is to investigate
the stability of the flat surface and the subsequent nonlinear
dynamics. 

We have seen just above that two different situations
may be distinguished according to
the presence (or not) of direct erosion by the wind. We will treat 
both situations and show that they lead to slight different dynamics.
We will deal first with the case where direct erosion by the wind
is present because it is the situation which has been treated
in \cite{Valance99}.\\

\begin{itemize}
\item {\bf  Presence of direct wind erosion}\\
This is the situation when the wind is not too strong. The
saltation curtain is not very  dense and the wind 
near the bed is strong enough to lift off some grains from the bed.
In this case, the exchange rate $\Gamma$ reads
\beq
\Gamma= \Gamma_{ej}^{imp} + \Gamma_{ej}^{wind} + \Gamma_{dep}
\pnt
\eeq
In that situation, the flat surface is found to be always
unstable. 
As soon as the wind is strong enough to maintain saltation 
and therefore induces reptation motion
(i.e., $\alpha_0 \neq 0$),
the surface is intrinsically unstable. In the 
situation where $\alpha_0/V$ is smaller than unity (which
is expected for low saltation flux; cf \cite{Valance99}) 
the dispersion relation in long wavelength limit,
is given by
\beq
\omega= \gamma_0 \left[ (\alpha_1 +\gamma_1)(\alpha_0/V) 
l_0^2 q^2 - l_0^3 l_c q^4 \right] 
+i \gamma_0 l_0^2 l_c q^3
\kom
\eeq
where $l_0= V/\gamma_0$ and $l_c=\beta_2/V$. 
$l_0$ is the reptating length for a flat bed while $l_c$ 
(which as the dimension of a length) plays the role
of a cut-off length preventing the surface from arbitrary small
wawelength deformation.
One clearly notes that the flat interface
destabilizes as soon as $\alpha_0$ is non zero. The most dangerous mode 
is given by
$\lambda_{max}= 2\pi \sqrt{l_0 l_c} / \sqrt{\varepsilon (\alpha_1+\gamma_1)}$
(where we set $\varepsilon= \alpha_0/V$). 
One can note that the most dangerous mode does not vary linearly with
the reptation length (as in the Anderson model) but is  given  by the
geometrical average between the reptation length $l_0$ and $l_c$.
This is a slight difference with the Anderson model. Unfortunately
the field observations and data from wind tunnel experiments do not allow us
to discriminate between these two descriptions.

In order to investigate the subsequent
development of the instability, the non-linear terms neglected
in the linear analysis should be taken into account. 
To do this, a non-linear analysis  is needed. 
By means of a multi-scale analysis,
it is possible to perform a weakly non-linear development
in the vicinity of the instability threshold
(i.e., $\varepsilon=\alpha_0/V \ll 1$).
We will not
expose the strategy of this analysis here (a detail
presentation can be found in \cite{Valance99}) but just
give the final outcome.
After some algebra, the non-linear analysis
yields an evolution equation for the bed profile which reads
\beq
\frac{\partial h}{\partial t} =  
f_1 \partial_{xx} h + f_2 \partial_{xxx} h + f_3 \partial_{xxxx}h
+f_{12} \partial_{xx}( h_x^2)  \label{rippleq1} \pnt
\eeq
$f_1= -l_0^2 (\alpha_1+\gamma_1) \varepsilon$,
$f_2=l_0 l_c$, $f_3=-l_0^3 l_c$  and 
$f_{12}=-l_0^2 l_c \gamma_1$.
We can note that the leading non-linear term is
of the form $\partial_{xx}( h_x^2)$.  
Using arguments based on symmetries and conservation laws 
(as to be seen in section 5), we would have expected a non-linearity
of the form $\partial_{x} (h_x^2)$. This term does not appear here. 
We may thus wonder whether it is fortuitous or not. It turns out that it is
an accident here because in the present situation
this term is of higher order
since it is multiplied by $\alpha_0$ (see eq. \ref{ej1}) 
which scales here as $\varepsilon$. We will see that in the next
case where there is no direct wind erosion, 
$\partial_{x} (h_x^2)$ is the leading order non-linear term.

\item {\bf Absence of direct wind erosion}\\
This situation occurs when the wind is relatively strong such
that the saltation curtain is dense enough to screen
the wind near the bead. This question was  not discussed previously
and constitutes an interesting point for the comparison with the
symmetry 
arguments developed later.
In this case, the erosion rate due to wind entrainment is
neglected so that the exchange rate $\Gamma$ is given by: 
\beq
\Gamma= \Gamma_{ej}^{imp} + \Gamma_{dep}
\eeq
A linear stability analysis teaches us that the flat bed surface
is unstable above a certain threshold defined by 
$\varepsilon=(\alpha_1 -\gamma_1) =0$. Indeed the growth
of a perturbation of the form $e^{iqx+\omega t}$ is given by
\beq
\omega \simeq \frac{\gamma_0 \alpha_0}{V} \left[ \varepsilon l_0^2 q^2  
-il_c l_0^2 q^3  
-l_c l_0^3 q^4 \right]
\eeq
where $l_c$ is now defined by $l_c=\alpha_2+ \gamma_2$.
The surface is unstable for $\varepsilon>0$ (i.e.,  $\alpha_1> \gamma_1$).
Since $\alpha_1=\cot \alpha$ (we recall that $\alpha$ is
the incident angle of the saltating grains), there exists
a critical incident angle $\alpha_c$ below which the flat bed is
unstable. In other words, grazing impact angle favours the ripple
instability. The most dangerous mode $q_{max}$ which is expected
to give the order of magnitude of the ripple wavelength
is easily estimated:
$ \lambda_{max}=2\pi/q_{max}= 2\pi \sqrt{l_o l_c}/\sqrt{\varepsilon}$.
Here again it is the geometrical
average between the saltation length $l_0$ and $l_c$.

A weakly non-linear analysis in the vicinity
of the threshold instability (i.e., $\varepsilon=\alpha_1-\gamma_1 \ll 1$)
can be performed following the same lines
as in the previous case. 
The calculation yields
\beq
\frac{\partial h}{\partial t}= 
f_1 \partial_{xx} h + f_2 \partial_{xxx} h + f_3 \partial_{xxxx}h  
+ f_{11}  \partial_x( h_x^2)  + f_{12} \partial_{xx} (   h_x^2) 
\label{rippleq2}
\eeq 
where $f_1= -l_0^2 \varepsilon$, 
$f_2=l_0 (\alpha_2+\gamma_2) \varepsilon^{-1/2}$,
$f_3=-l_0 (\alpha_2+\gamma_2) \varepsilon^{-1/2}$,
$f_{11}=l_0 \alpha_3$ and 
$f_{12}=(\alpha_4 -\alpha_3 l_0 - \gamma_1 l_c) \varepsilon^{1/2}$.  
The leading non-linear term is 
$\partial_x( h_x^2)$, that is the non-linear term
expected from the symmetries as to be seen below. The non-linear term 
coming to next order is
$\partial_{xx} ( h_x^2)$  and we will see
that this second non-linearity is crucial to stabilize the linear
growth of the structure. \\

\end{itemize}

\def\beq#1{\begin{equation}\label{eq:#1}}
\def\eeq{\end{equation}}
\def\beqa#1{\begin{eqnarray}\label{eq:#1}}
\def\eeqa{\end{eqnarray}}
\def\vt{v_{\rm t}}
\def\vn{v_{\rm n}}
\def\qc{q_{\rm c}}
\def\lc{\lambda_{\rm c}}
\def\qco{q_{\rm cut-off}}
\def\lco{\lambda_{\rm cut-off}}
\def\ds{{\rm d}s}
\def\avg#1{\left\langle #1 \right\rangle}
\def\dt{\partial_t}
\def\w{\omega}
\def\inf{\infty}
\def\e{\epsilon}
\def\Re{{\rm Re}~}
\def\Im{{\rm Im}~}
\def\EQ#1{Eq.~(\ref{eq:#1})}
\def\FIG#1{Fig.~\ref{#1}}
\def\REF#1{~\cite{#1}}

\section{Non-linear ripple dynamics}

In the previous sections we have seen how can a mathematical model
be constructed for ripple formation.
It is  natural to
ask whether there is a simple explanation why
Eq.~(\ref{rippleq2}) is the governing
equation of ripple formation.
It turns out that evoking only geometrical and conservation 
considerations it is possible to predict the form of the equation
including the leading non-linear terms.  The power of this
approach, as was used in a more general context recently\REF{Csahok99a},
is that it is model-independent, and that it can provide
very general ingredients for the appearance of a nonlinearity,
as we shall comment below. In particular, it will, appear, for example,
that
though the nonlinearity $\partial _{xx}(h_x^2)$ is compatible
with symmetries and conservation, it should not be present
if the system where not anisotropic! (here the wind).  

\subsection{General approach}
To start with let us consider an arbitrary (not self crossing)
curved front in one dimension. It represents the sand--air
front parametrized by an intrinsic variable $\alpha$ ($0<\alpha<1$).
In a coordinate system independent representation
the front can be characterized (up to a rotation and displacement)
by its curvature $\kappa$ as a function of the arclength $s$.
It is conceptually important to make a clear distinction
between $\alpha$ and $s$. For example, $\alpha=1$ always corresponds to the end
of the curve while the  arclength  coordinate of the end  (i.e., the total
length of the curve) can change. It is therefore not equivalent to work
at constant $\alpha$ or at constant $s$.

We are interested in deriving a general form of
evolution equation for the front.
More precisely we are seeking the equation of evolution of
the curvature.  From geometrical considerations \REF{Csahok99a} we obtain
the following equation
\beq{cevol}
\kappa_t\big\vert_s =
-\left[{\partial^2\over \partial s^2} +\kappa^2 \right]v_n
   -{\partial \kappa\over \partial s}\int _0^s ds' \kappa v_n,
\eeq
where $\vn$ denotes the normal component of the
local velocity of the surface. This  is a general equation
which holds for any front.

The normal velocity ($\vn$) contains the physics of the evolution
process of the surface.
Since $\vn$ is a coordinate system independent quantity
(i.e., a scalar)
it must be
a function of the curvature and its derivatives with respect to
the arclength (that are also scalars).
The knowledge of $\vn(\kappa)$ allows
to obtain the dynamics of the front from \EQ{cevol}.
In the general case, however, it is possible only by
numerical integration of the equation. Note also that 
the above equation is very appropriate for numerical treatment
in an intrinsic representation.

In our particular case we restrict ourselves to slightly curved fronts
that will allow to derive the evolution equation in a closed form.
There is a privileged direction in the ripple formation process as
the \mbox{$x\leftrightarrow -x$} symmetry is broken due to the wind.
Therefore $\vn$ may contain explicit dependence on the local slope
$\theta$ of the surface
\beq{vn1}
\vn = \vn(\theta, \kappa,\kappa_s,\dots).
\eeq
Since $\kappa=\theta_s$, we can reformulate \EQ{vn1} as
\beq{vn2}
\vn = \vn(\theta, \theta_s,\theta_{ss},\dots).
\eeq

The concrete
choice of this dependence is restricted by the mass conservation law for
the sand
\beq{conss}
\oint \vn \ds = 0.
\eeq
This condition eliminates, for example, a choice like $\vn \sim \kappa^2$.
If the evolution process can be considered as local, as in the case when
the reptation length is much smaller than the ripple wavelength,
we can write \EQ{vn2} as
\beq{noname1}
\vn = \frac{\partial}{\partial s} F(\theta, \theta_s,\theta_{ss},\dots),
\eeq
where $F$ is an arbitrary (but smooth) function of its arguments.
Expanding $F$ around $\theta=0$ (straight front)
we obtain
\beq{vn}
\vn = \frac{\partial}{\partial s} \left(
f_1 \theta + f_2 \theta_s + \dots
+\frac{1}{2}f_{11}\theta^2 + f_{12}\theta\theta_s+
\frac{1}{2}f_{22} \theta_s^2 + \dots
+\frac{1}{6}f_{333}\theta^3 + \dots \right)
\eeq

The assumption of a slightly curved front
(the height of the ripples $H$ is always much smaller than their
wavelength $\lambda$ in the experiments,
$\partial h/\partial x \sim H/\lambda \ll 1$)
allows us to describe the
front by a more natural parameter: its height $h(x,t)$ with respect
to the initial state (\FIG{explain}).
The inclination $\theta$ and the curvature $\kappa$ can be expressed
in terms of $h$ in leading order  as $h_x$ and $h_{xx}$, respectively.
Substituting \EQ{vn} into \EQ{cevol} and keeping only the lowest order
{\it linear} terms
results in 
\beq{linhevol}
h_t = f_1 h_{xx} + f_2 h_{xxx} + f_3 h_{xxxx}. % + (h_x^2)_{xx}.
\eeq
The first term on the r.h.s corresponds to a sum of diffusion
in the fluidized upper layer
driven by gravity
(i.e., rolling in the model of Hoyle\REF{Hoyle97}) and
the effect of erosion by the wind. The
third term represents surface diffusion that
comes from an effective
surface tension (also related to the property of the fluidized layer)
so its prefactor $f_3$ is considered
to be always negative (stabilizing).
If the prefactor of the first term
$f_1>0$ then the flat interface ($h=0$) is stable.
If, on the other hand, $f_1<0$ then the flat interface becomes unstable
against ripple formation.
The second term in \EQ{linhevol} is a propagative term that is responsible
for the drift of the emerging pattern.
In fact, a term proportional to $h_x$ is
also acceptable in \EQ{linhevol} but it can be easily eliminated by
a Galilean transformation.

\subsection{Wavelength selection at short times}

In the previous section we have seen that analyzing the physical
processes during sand ripple evolution one finds the prefactor of $h_{xx}$
can change its sign and become negative in case of sufficiently strong winds.
This leads to the appearance of a range of linearly unstable modes
for $f_1<0$.
The linear dispersion relation of fluctuations around $h=0$ is
\beq{w}
\w(q) = - f_1 q^2 -i f_2 q^3 + f_3 q^4,
\eeq
where $q$ is the wavenumber ($h\sim e^{\w t + iqx}$).
The growth rate of fluctuations
is determined by the real part of the dispersion relation
($-f_1 q^2 + f_3 q^4$) while the imaginary part describes the drift
properties.
The wavenumber of the linearly most unstable mode is $\qc=\sqrt{f_1/(2f_3)}$
(note that $f_1, f_3 <0$)
that gives the typical ripple wavelength ($\lc=2\pi/\qc$)
which is observed shortly after their appearance.

To proceed, we have to identify which {\it non-linear} terms have
the most important contribution to \EQ{linhevol}.
Consider a ripple structure of wavelength $\lambda$.
Its amplitude ($H$) will be infinitesimal at $t=0$ and then
grow exponentially due the linear instability.
We can write the typical height as $H\sim \lambda^a$, where
$a=a(t)<0$ is an increasing function of time.
The order of magnitude of the terms in  \EQ{vn} can be easily evaluated:

\begin{center}
\begin{tabular}{ccccc}
$\theta_{ss}$	&$\sim$& $h_{xxxx}$	&$\sim$& $\lambda^{a-3}$ \\
$\theta^2$	&$\sim$& $(h_x^2)_x$	&$\sim$& $\lambda^{2a-2}$ \\
$\theta\theta_s$	&$\sim$& $(h_x^2)_{xx}$	&$\sim$& $\lambda^{2a-3}$ \\
$\theta^3$	&$\sim$& $(h_x^3)_x$	&$\sim$& $\lambda^{2a-3}$ \\
\end{tabular}
\end{center}

The dominant terms are the ones with largest exponent. That is when
$a$ is a large negative number ($t\simeq 0$) then all the non-linear
terms are negligible compared to the linear terms as expected.
The first non-linearity that becomes significant is  $(h_x^2)_x$.
%Analyzing \EQ{vn} taking into account the typical length scale $\lc$
%and amplitude of the pattern
%one can easily find that the lowest order
%non-linear term is $f_{22} (h_x^2)_x$.
In fact, this term contains an odd number of spatial derivatives
and therefore it gives contribution only
to the imaginary part of $\w(q)$.
Consequently,  it can not lead to a development
of a finite height structure although it modifies the drift properties.
We disregard for the moment the term $(h_x^2)_x$ to which we
will come back later.
The next lowest order term is $(h_x^2)_{xx}$ which leads indeed to
saturation.
We dispose of three physical scales in the problem
(time, length, and height) and \EQ{linhevol} extended with the before
mentioned non-linear term contains four
relations between the prefactors of the terms.
Thus by appropriate rescaling\footnote{With an equation
having 5 terms, we have 4 independent coefficients. Rescaling
space, time and the height, we can absorb 3 of them so that
the equation can be reduced to a one parameter one.} variables the full
equation can be reduced to
the following single parameter
non-linear evolution equation for the ripple height
\beq{hevol1}
h_t =  -h_{xx} + \nu h_{xxx} +  h_{xxxx} + (h_x^2)_{xx}.
\eeq
The sign of the non-linear term is taken to be positive.
Choosing negative sign would be equivalent simply to a
$h\leftrightarrow -h$ transformation of the original.
Since \EQ{hevol1} has no up--down symmetry simply by
inspecting the form of the ripples one can decide if it is
the positive or the negative sign that corresponds to the physical
situation (\FIG{explain}). Apparently, it is the positive sign
that is appropriate
in case of both aeolian and under water ripples.

\subsection{Amplitude expansion}

To analyze the properties of \EQ{hevol1} let us consider
the stationary ($h_t=0$) solutions of \EQ{hevol1} with spatial
period $L$ and for the moment with $\nu=0$.

\begin{figure}
\begin{center}
\includegraphics[height=9cm]{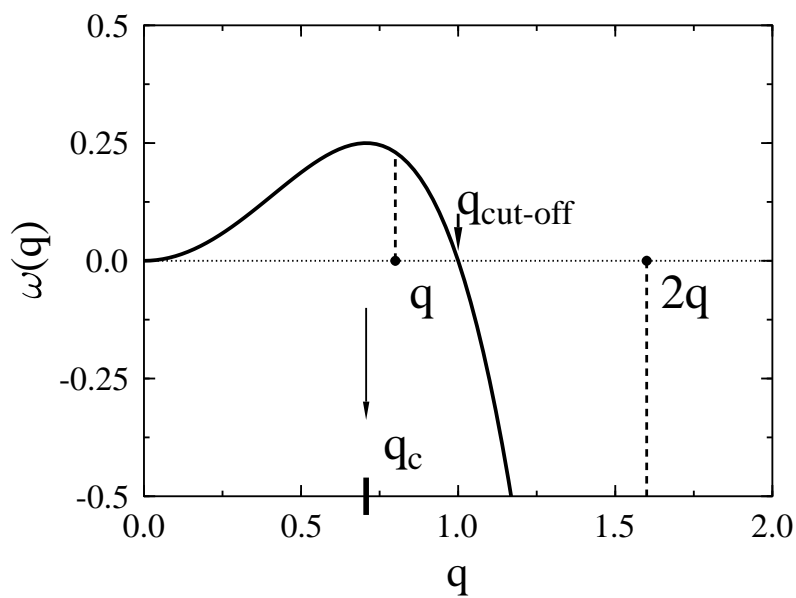}
\end{center}
\caption{Dispersion relation $\w (q)$. The points $q$ and $2q$
show the two modes used in deriving \EQ{A1}. Note that
$\w (q)>0$ while $\w (2q)<0$.}
\label{disp1}
\end{figure}

%::: $A_k$ :::
The first remark that has to be made is that
the instability of the planar solution can manifest itself
only if the lateral size of the system $L$
is larger than $\lco=2\pi$. This feature is due to the fact
that the largest wavenumber in the unstable band
is given by $2\pi/L$.
Thus, if the size of system is too small then all possible Fourier
modes will be stable (\FIG{disp1}).
In order to find the amplitude of the developed pattern
we re-write \EQ{hevol1} in Fourier space
\beq{aevol}
A_n = \w(nq) A_n + q^4 n^2 \sum_{m=-\inf}^{\inf} m(n-m) A_m A_{n-m},
\eeq
where $A_n$ is the amplitude of the Fourier mode with wave number $nq$,
i.e., $h(x)= \sum_{n=-\inf}^{\inf} A_n e^{inqx}$.
The amplitudes are subject to the restriction $A_n= \bar{A}_{-n}$
(since $h(x)$ is a purely real function) and $A_0\equiv 0$ (since
we impose $\int_{-\inf}^{\inf}h(x) dx=0$). 
%The latter propriety is equivalent to restricting the sum

If $L\simeq\lco$ then
only the longest wavelength mode ($A_1$) is active (i.e., instable).
The first harmonic ($A_2$) is inherently stable but since it is
coupled through the non-linear term to the leading mode,
it will be non-zero.
The higher harmonics can be safely neglected as their amplitude will
be exponentially small compared to $A_2$.
We take into account only the first two modes and in addition
we can assume that $A_2$ varies much faster
than $A_1$, and hence $A_2$ is {\it adiabatically slaved} to $A_1$.
Solving the resulting set of two equations we find for the leading
amplitude
\beq{A1}
A_1^2 = -\frac{\w(q)\w(2q)}{16 q^8}, %(1-q^2)(4q^2-1)/(4q^4)
\eeq
where $q=2\pi/L$.
In order to have a solution for $A_1$ the r.h.s of \EQ{A1}
has to evaluate to a non-negative real number.
Therefore two conditions has to be satisfied:
(i) $-\w(q)\w(2q) > 0$ and (ii) $\Im (\w(q)\w(2q))$=0.
Since $\w(q)>0$ and $\w(2q)<0$, and both are real, the
two conditions are met.
It is convenient to choose $A_1$ to be real
(\EQ{A1} fixes only the magnitude of $A_1$) and then
it scales as $A_1\sim (\qco-q)^{1/2}$ where $\qco=1$.
The approximation that only one mode is active breaks down far
from the threshold ($L\gg \lco$ or $q\ll\qco$).
Indeed, for $q=\frac{1}{2}\qco$ \EQ{A1} gives
a zero amplitude that is obviously not correct.

Far from the threshold
we consider that the wavelength of the pattern
is very large (i.e., $q\ll\qco$) and thus take approximately
$\w(nq)\simeq n^2 q^2$.
To remove the dependence on $q$ from \EQ{aevol} we
look for a solution in form of
$A_n\sim q^{-2}$.
After some algebra the amplitude of the $n$th mode is found to be
\beq{An}
A_n = \frac{1}{2 n^2 q^2}.
\eeq
This relation is expected to be valid only if $nq\ll 1$.
Figure \ref{ampliq2}
shows that \EQ{An} reproduces very well the direct numerical
solution of \EQ{aevol}.

\subsection{Coarsening}

There is a subtle issue that is worth emphasizing.
A stationary solution $h_L(x)$ with period $L$ will be a stationary 
in a box of $2L$ too. That is -- if $L$ is large enough --
there can be multiple solutions with periods that are divisors of $L$.
Which one of these is stable? We find by numerical stability analysis
that the solution with the longest period is the stable one.
This means that during the temporal evolution from a planar front
first the fastest growing mode appears and then the structure 
gradually coarsens to reach the final state of one huge ridge.

The width $w^2=\avg{h^2}$ of the pattern evolves in time as
\beqa{wevol}
\frac{1}{2}\dt w^2 &=& \avg{h \dt h} =
\avg{h\left(-h_{xx}-h_{xxxx}+(h_x^2)_{xx}\right)}  \nonumber\\
& = & \avg{h_x^2} - \avg{h_{xx}^2} + \avg{h_{xx} h_x^2}.
\eeqa
The last term is zero since it is a full derivative
with respect to $x$.
The growth of the width is due to the first term, the second
one being negligible for later times.
The typical wavelength and slope scale with time
as $\lambda\sim t^{1/z}$ and $\theta\sim t^\alpha$, respectively.
Using \EQ{An} we find that the amplitude $H$ of the structure
behaves as
\beq{Alambda}
H\sim A_1 \sim \lambda^2.
\eeq
On the other hand $H$ can be approximated as
\beq{noname2}
H \sim \lambda\theta.
\eeq
Combining these two relations with the scaling
of $\lambda$ and $\theta$
gives the exponent relation
\beq{exprelation}
\alpha = 1/z.
\eeq
The width of the interface scales as $w\sim H$ and
thus the order of the terms of \EQ{wevol} is written as
\beq{orders}
O(t^{4/z-1}) = O(t^{2/z}) - O(1).
\eeq
We see that the growth is dominated by the first term on the r.h.s
that corresponds to the unstable linear term.
By equating the exponents on the two sides of \EQ{orders} we
obtain for the coarsening exponent 
\beq{zhalf}
1/z= 1/2,
\eeq
in accord with the results
of numerical simulation  (\FIG{lambdevol}).

\begin{figure}
\begin{center}
\includegraphics[height=9cm]{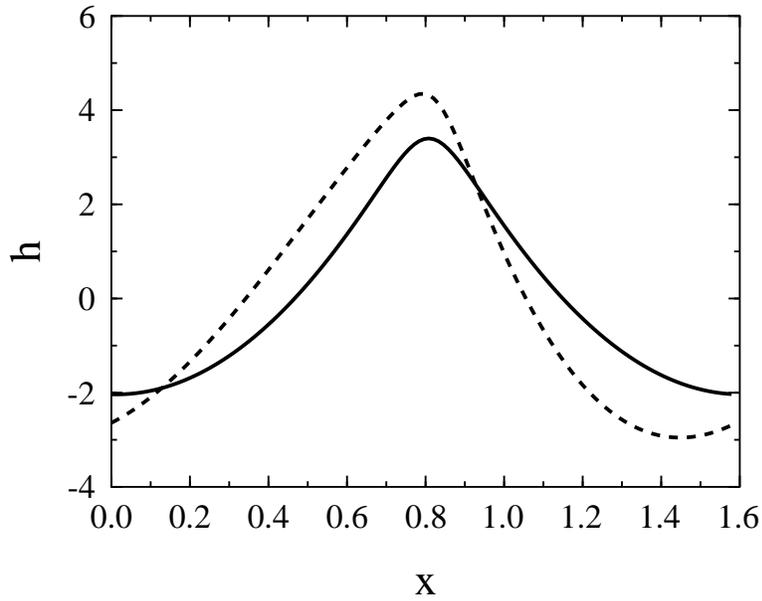}
\end{center}
\caption{Front shape (solid line) for $\nu=0$ and (dashed line) for $\nu=3$.}
\label{sample}
\end{figure}

\begin{figure}
\begin{center}
\includegraphics[height=9cm]{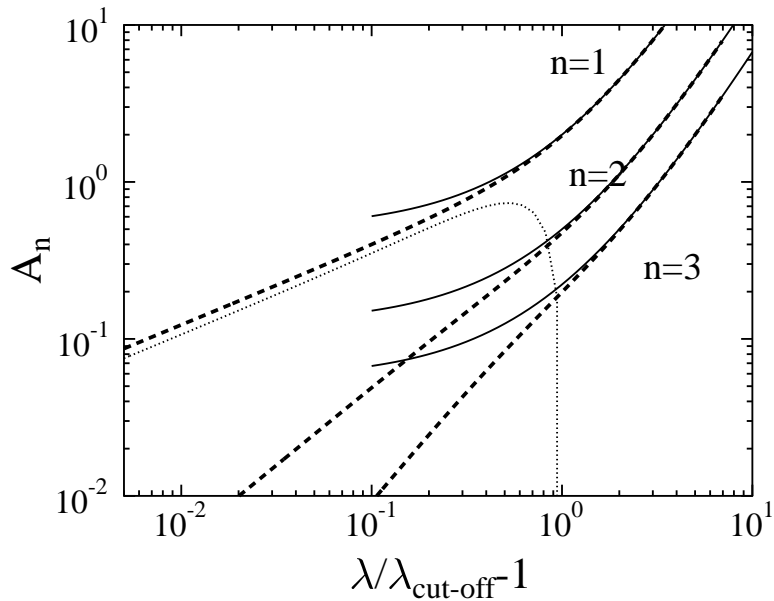}
\end{center}
\caption{Amplitudes of the leading Fourier modes of the ripple
structure. Thick dashed line shows numerical solution of \EQ{aevol},
thin line shows the far-from-threshold result (\EQ{An}), and
dotted line corresponds to the two mode approximation (\EQ{A1}).}
\label{ampliq2}
\end{figure}

\begin{figure}
\begin{center}
\includegraphics[height=9cm]{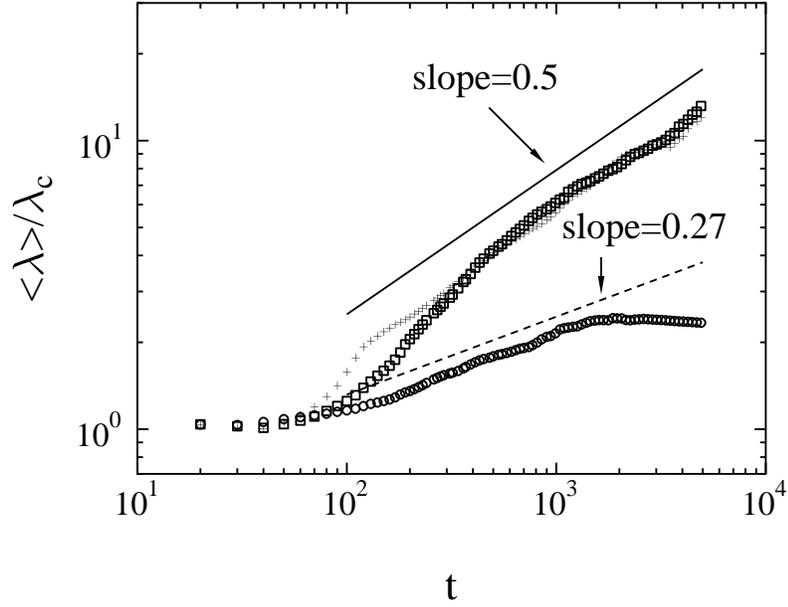}
\end{center}
\caption{
Temporal evolution of the average wavelength integrating \EQ{hevol1}
(pluses) for $\nu=0$, (boxes) for $\nu=1$, and (circles) \EQ{hevol2}
with $\nu=\mu=\eta=1$.
}
\label{lambdevol}
\end{figure}

\begin{figure}
\begin{center}
\includegraphics[height=9cm]{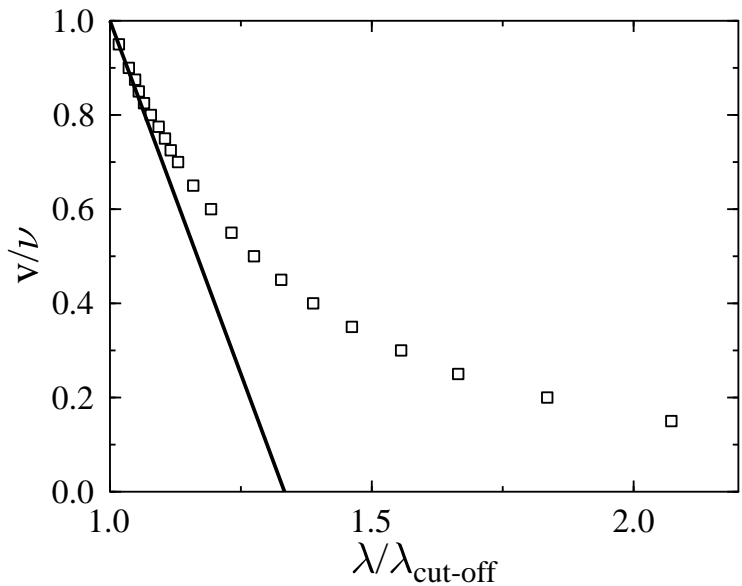}
\end{center}
\caption{Ripple drift velocity $v$ as a function of the ripple
wavelength. (Boxes) numerical results from integrating \EQ{hevol1}
with $\nu=1$ and (solid line) analytical result for
$\lambda/\lc\simeq 1$ \EQ{vdrift}.
}
\label{drift}
\end{figure}

%::: drift :::
For $\nu>0$ the pattern loses its $x \leftrightarrow -x$ symmetry
(\FIG{sample})
and drifts sideways. We have measured the drift velocity
numerically (\FIG{drift})
and close to the threshold 
it compares well with the results of calculation around the threshold
\beq{vdrift}
v = \nu - 3\nu (\lambda/\lco-1) + O( (\lambda/\lco-1)^2 )
\eeq
This equation results from the requirement (see \EQ{A1}) that
$\Im \w(q)\w(2q)=0$.
The imaginary contribution originating from
the $\nu h_{xxx}$ term can be compensated by a purely propagative
term $v h_x$ that fixes the value of the drift $v$.
We find that the coarsening law for drifting patterns does not
change, the scaling $t^{1/2}$ is observed (\FIG{lambdevol}).
This is not surprising since considering a $h_{xxx}$ term
in \EQ{wevol} leads to a zero contribution as
$\langle h h_{xxx}\rangle = -\langle h_x h_{xx}\rangle = 0$.
%
%!!!!!!!!!!!!!!!!!!
% reparler de terme (h_x^2)_x
%

\subsection{Higher order non-linearities}

%::: $(h_x^3)_x$ :::
As it can be easily seen from \EQ{A1}
the amplitude of the basic Fourier mode ($A_1$) for the solutions
of \EQ{hevol1} grows as $L^2$. That is the typical slope
of the structure ($A_1/L$) increases indefinitely as the wavelength
increases during coarsening.
Another way to view this feature is realizing that \EQ{hevol1} possesses
a parabolic particular solution for $\nu=0$
of form  $h(x)=h_0-\frac{1}{4}x^2$.
Introducing the next order non-linear term 
($(h_x^3)_x$) limits the growth of the amplitude 
to be of the order of $L$ and thus imposes a finite slope.
In fact, in the derivation of \EQ{hevol1} we have supposed that
the slope is small and therefore that equation is only valid at the birth
of the ripple structure. For later times higher order non-linearities
start to play an important role and thus the evolution equation
changes to
\beq{hevol2}
h_t =  -h_{xx} + \nu h_{xxx} +  h_{xxxx} +
	\mu(h_x^2)_{xx} + \eta(h_x^3)_x.
\eeq
Here we have introduced the parameters $\mu$ and $\eta$ to control
the relative importance of the two non-linear terms.

%::: discuss the effect of (h_x^3)_x :::
If we set $\mu=0$ then the $h\leftrightarrow -h$ symmetry of the
equation is restored. As a consequence at later stages of ripple
development when the second non-linearity becomes dominant
the shape of ripples will become more triangular.
If in addition $\nu=0$ 
then the $x\leftrightarrow -x$ symmetry is restored and
the system becomes variational.
In this limit
\EQ{hevol2} reduces to
the noiseless conserved Cahn-Hilliard equation \REF{Politi98}.
The coarsening takes place very slowly, the typical length scale
grows as
\beq{logcoars}
\lambda\sim \ln t.
\eeq
By setting non-zero $\nu$ and $\mu$, that is imposing a drift
($h_{xxx}$ term) and
re-introducing the leading non-linearity 
($(h_x^2)_{xx}$ term) we observe an effective
scaling of the wavelength over almost one decade. The exponent
is found to be close to $1/4$. But the coarsening process stops
after some time, meaning that the surface regains its stability.
This effect can be attributed to the stabilizing nature of
the $h_{xxx}$ term: it introduces a wavelength dependent drift
velocity for the perturbations and thus diminishes their coherence
leading to effective stabilization.
The non-linear term $(h_x^2)_{xx}$, on the other hand, acts on the
direction of destabilizing the surface and accelerates the coarsening.
Far from the threshold, however, the importance of this term becomes
negligible and the second non-linearity dominates the dynamics.
Since in the case  $\nu=0$ considered above the coarsening was logarithmic,
so in a sense marginal, introducing a stabilizing term can lead to
an eventual stopping of the coarsening process.
This was not the case with only the leading non-linearity as we have
seen above: there the scaling is not affected by the extra linear term.

\begin{figure}
\begin{center}
\includegraphics[angle= -90,scale=0.5]{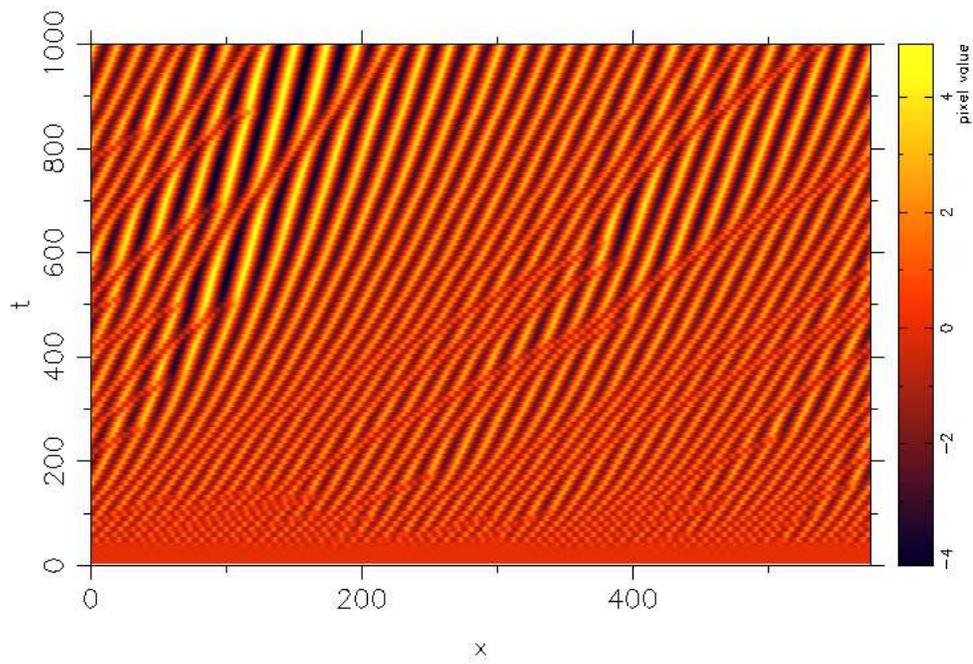}
\end{center}
\caption{
Temporal evolution of the ripples. Note the coarsening and the
slowing down of the drift.
}
\label{rippevol}
\end{figure}

The numerical results presented in the figures
has been obtained by
the integration of \EQ{hevol1} and \EQ{hevol2}
using a pseudo-spectral method
with $\Delta x= 0.1$ and $\Delta t= 10^{-5}$ in a system of size
$L\sim 30\lco$.
Figure \ref{rippevol} shows the temporal evolution of the structure
corresponding to the  case $\nu=\mu=\eta=1$.
The simulation has been started from a small amplitude
random initial condition. Soon a rather regular pattern appears with
wavenumber corresponding to the linearly most unstable mode.
The structure contains defects that will trigger the coarsening process:
at the end of the simulation the typical wavelength has been doubled.
The slowing down of the drift with growing wavelength predicted by
\EQ{vdrift} is also clearly observable.

\section{Conclusion and discussion}

Based on general arguments and observations we have
adopted the notion of locality for sand ripple formation.
We have then presented general symmetry and conservation
considerations to show how a  model-independent
nonlinear equation for sand ripples can be derived. That equation
takes the form 

\beq{hevol3}
h_t =  -h_{xx} + \nu h_{xxx} +  h_{xxxx} +
(h_x^2)_x+        \mu(h_x^2)_{xx} + \eta(h_x^3)_x.
        \eeq
        The first nonlinearity $(h_x^2)_x$ contributes to drift and 
        is not able to saturate the linear growth. So the first
        efficient nonlinearity is $\mu(h_x^2)_{xx}$. At short time
        an ordered ripple structure emerges with a wavelength 
        close to that corresponding to the linearly fastest growing mode.
Later a coarsening process takes place; With $\eta=0$ coarsening 
continues indefinitely, until one huge dune is reached. The 
 coarsening
 is quantified in terms  of a dynamical exponent defined  in relation
 to the increase of 
  the mean wavelength,  $\lambda\sim t^{1/z}$.
 We find both analytically and numerically $z=2$. We have
 also identified that the slope increases in that case
 without bound when increasing the system size. Thus higher
 nonlinear terms must become decisive. We have included the
 next nonlinear term $(h_x^3)_x$ which has led to
 a saturation of the slope (both the height and the wavelength
 scale in the same manner as a function of the system size).
 Though at short time this nonlinear term is irrelevant, it
 dominates dynamics at longer times. Here again we observe a 
 coarsening but with a  smaller exponent $1/z=1/4$. The coarsening seems
 to stop after a certain stage, typically when the wavelength 
 is about twice of the fastest growing mode.
 We have found that as the ripple structure forms, it drifts
 sideways. The drift occurs with dispersion (the drift velocity depends
 on wavelength). The first term which is responsible 
 for the drift is the one proportional to $h_{xxx}$ (note
 that there is another linear term $h_x$ which provides
 a phase velocity that can be absorbed in $h_t$ via a Galilean
 transformation). The nonlinear term $(h_x^2)_x$ though it does
 not saturate the linear growth it contributes significantly
 to drift. In particular it can also lead to  a drift 
 opposite to the wind. This happens in particular when
all three nonlinearities of \EQ{hevol3} are present but
$\nu=0$.

The hydrodynamical model captures the full nonlinear equation
written above and provides an encouraging physical
basis for the derivation of the ripples equation from 
physical ingredients. Unlike
symmetry and conservation laws, the explicit physical
model relates the coefficients to the underlying (phenomenological)
physical parameters
and provides the physical explanation for the initiation
of the instability. That instability is also present 
in a very transparent and general picture in the Anderson's model.

These two views (explicit phenomenological model and symmetries)
have aided in identifying some general picture of the form of the
ripple evolution equation. A numerical study, though not exhaustive,
has allowed extraction of some general results. It will be of great importance
in future studies to quantify experimentally the coarsening process
and to identify whether or not the  coarsening stops
or rather would continue without bounds if energy continues
to be injected. This step will be vital  to guide further
theoretical development.

\newpage

\end{document}